\documentstyle[prd,aps,psfig,subfigure,floats,preprint,tighten]{revtex}

\newcommand{\CL}{{\cal L}}

\newcommand{\CO}{{\cal O}}
\newcommand{\CP}{{\cal P}}

\newcommand{\bea}{\begin{eqnarray}}  \newcommand{\eea}{\end{eqnarray}}
\newcommand{\beq}{\begin{equation}}  \newcommand{\eeq}{\end{equation}}
\newcommand{\non}{\nonumber}  
\newcommand{\lmk}{\left(}  \newcommand{\rmk}{\right)}
\newcommand{\lkk}{\left[}  \newcommand{\rkk}{\right]}
\newcommand{\lhk}{\left \{ }  \newcommand{\rhk}{\right \} }
\newcommand{\del}{\partial}  
\newcommand{\vect}[1]{\mbox{\boldmath${#1}$}}
\newcommand{\bib}{\bibitem} 
\newcommand{\la}{\left\langle} \newcommand{\ra}{\right\rangle}

\def\IB#1#2#3{{\bf #1}, #2 (19#3)}
\def\IBD#1#2#3{{\bf D#1}, #2 (19#3)}
\def\NPB#1#2#3{Nucl. Phys. {\bf B#1}, #2 (19#3)}

\def\PLB#1#2#3{Phys. Lett. {\bf B#1}, #2 (19#3)}
\def\PLBold#1#2#3{Phys. Lett. {\bf#1B}, #2 (19#3)}

\def\PRD#1#2#3{Phys. Rev. {\bf D#1}, #2 (19#3)}

\def\PRL#1#2#3{Phys. Rev. Lett. {\bf#1}, #2 (19#3)}
\def\PRT#1#2#3{Phys. Rep. {\bf#1}, #2 (19#3)}

\def\CP#1#2#3{Computers in Physics {\bf #1}, #2 (19#3)}

\def\PRS#1#2#3{Proc. Roy. Soc. {\bf A#1}, #2 (19#3)}

\def\JP#1#2#3{J. Phys. {\bf A#1}, #2 (19#3)}

\begin{document}


\begin{center}
{\Large \bf Scaling Property of the global string in the radiation
  dominated universe \\}
\vskip 1cm
{\large Masahide Yamaguchi} \\
\vskip 0.5cm
{\large \em Department of Physics, University of Tokyo} \\
\vskip 0.1cm
{\large \em Tokyo 113-0033, Japan}
\end{center}


\begin{abstract}
  We investigate the evolution of the global string network in the
  radiation dominated universe by use of numerical simulations in 3+1
  dimensions. We find that the global string network settles down to
  the scaling regime where the energy density of global strings,
  $\rho_{s}$, is given by $\rho_{s} = \xi \mu / t^2$ with $\mu$ the
  string tension per unit length and the scaling parameter, $\xi \sim
  (0.9-1.3)$, irrespective of the cosmic time. We also find that the
  loop distribution function can be fitted with that predicted by the
  so-called one scale model. Concretely, the number density,
  $n_{l}(t)$, of the loop with the length, $l$, is given by $n_{l}(t)
  = \nu/[t^{3/2} (l + \kappa t)^{5/2}]$ where $\nu \sim 0.0865$ and
  $\kappa$ is related with the Nambu-Goldstone(NG) boson radiation
  power from global strings, $P$, as $P = \kappa \mu$ with $\kappa
  \sim 0.535$. Therefore, the loop production function also scales and
  the typical scale of produced loops is nearly the horizon distance.
  Thus, the evolution of the global string network in the radiation
  dominated universe can be well described by the one scale model in
  contrast with that of the local string network.
\end{abstract}

\thispagestyle{empty} \setcounter{page}{0} \newpage
\setcounter{page}{1}

\section{Introduction}

\label{sec:introduction}

Cosmic strings could be formed in consequence of the cosmological
phase transition at the very early universe \cite{KIB}. They are
divided into global and local(gauged) strings according to the
property of the broken symmetry. Though they have similar properties,
such as they are line objects with the false vacuum energy,
intercommute at crossing \cite{SHE,MMMR} and so on, there are also
many differences between them. While a local string has two cores
comprised of a magnetic flux core and a scalar field core, a global
string has only a scalar field core. For local strings, the gradient
energy of the scalar field can be canceled out by the gauge field far
from the core so that the core is well localized and the vacuum energy
of the core is dominant. Therefore, the Nambu-Goto action is adequate
to follow the evolution of the local string network except at
crossing. On the other hand, there is no gauge field in the global
string model so that the total energy of global strings is dominated
by not the vacuum energy of the core but the gradient energy of the
scalar field, namely, the NG boson field. So, we need consider not
only the core but also the associated NG boson field and the coupling
between them in contrast with the case of local strings. Thus we must
use the Kalb-Ramond action instead of the Nambu-Goto action as an
effective action \cite{KR}.

The dynamics of the local string network has been examined by use of
the Nambu-Goto action. Kibble \cite{KIB2} first proposed the so-called
one scale model, where the behavior of the system can be characterized
by only a parameter, namely, the scale length, $L$, and an unknown
loop-production function. He showed that either the local string
network goes into the scaling regime where the scale length, $L$,
grows with the horizon distance \cite{KIB,KV}, or, the universe
becomes string dominated. In the scaling regime, infinite strings
intercommute to produce closed loops so that at any time the number of
strings stretching across the horizon distance within each horizon
volume is almost constant and produced loops decay through radiating
gravitational waves \cite{VIL}. Bennett \cite{BEN} developed the
Kibble's one scale model and showed that unless most of produced loops
self-intersect and fragment into smaller loops with the typical length
smaller than $L$, the reconnection rate is large enough to prevent
scaling. Mitchell and Turok \cite{MT} studied the statistical
mechanics of the string network in the flat spacetime and found that
the equilibrium distribution of the string network is dominated by the
smallest loops allowed, which suggested that strings tend to break
into very small pieces in the expanding universe.  Albrecht and Turok
\cite{AT} modeled the network as the hot body radiation where loops
are radiated from the long strings and showed that the scaling
solution is inevitable. However, the application of the flat spacetime
statistical mechanics to the string dynamics in the expanding universe
is not necessarily justified. Thus, numerical simulations are
unavoidable to decide whether the local string network settles down to
the scaling regime or not.

Albrecht and Turok \cite{AT2} gave the first numerical investigation
for the scaling property. Later three groups \cite{AT,BB,AS} improved
numerical codes and all groups concluded that the large scale behavior
of the local string network goes into the scaling regime with the
scaling parameter $\xi$ equal to $(50\pm25)$ \cite{AT}, $(13\pm2.5)$
\cite{BB}, and $(16\pm4)$ \cite{AS} in the radiation-dominated era.
Though Albrecht and Turok \cite{AT} found that the loop distribution
function also scales, the higher resolution simulations performed by
the other two groups \cite{BB,AS} showed that long strings have
significant small structures and loops are typically produced at
scales much smaller than the horizon distance, which is close to the
cut-off scale. In response, Austin {\it et al}. \cite{ACK} proposed
the three scale model with $\xi$, the step length $\bar{\xi}$, and the
scale $\zeta$ describing the small scale structure. They found that
$\xi$ and $\bar{\xi}$ grow with the horizon distance but $\zeta$ begin
growing only when the gravitational back reaction becomes in effect
with $\zeta/\xi \sim 10^{-4}$.  However, Vincent {\it et al}
\cite{VHS} recently claimed from the flat spacetime simulations that
instead of loop production due to intercommutation, long strings
directly emit massive particles so that the dominant energy loss
mechanism of local strings is not gravitational radiation but particle
production.

Thus, though there is a consensus that the large scale structure of
the local string network obeys the scaling solution, the loop
production and the dominant energy loss mechanism are now in dispute.
This is mainly because inclusion of gravitational radiation to the
numerical simulations is impossible and gravitational radiation is so
weak that kinks live for a long time.

On the other hand, the evolution of the global string network has been
less studied. Its manifestation, however, is essential to estimation
of abundances of relic axions radiated from axionic strings
\cite{DA,BS2,flat}, which may be the cold dark matter. Also, scaling
property is indispensable for the scenario where global strings become
the seed of the structure formation of the universe and produce the
cosmic microwave background anisotropy \cite{PST}. Thus, it is
important to clarify the dynamics of the global string network, in
particular, whether it enters the scaling regime like the local string
network.

So far, the result for the local string network has been applied to
the global string network because global strings also intercommute
with the probability of the order unity \cite{SHE}. But global strings
have the associated Nambu-Goldstone bosons, which lead to long-range
forces between strings and become the dominant energy loss mechanism
of global strings. Therefore, instead of the Nambu-Goto action we have
to use the Kalb-Ramond action \cite{KR}, which is comprised of three
components, the Nambu-Goto action, the kinetic term of the associated
Nambu-Goldstone fields, and the coupling term between them. But the
Kalb-Ramond action has difficulty of logarithmic divergence due to
self-energy of the string. Dabholkar and Quashnock \cite{DQ} solved
this difficulty by the similar prescription as given by Dirac in the
electromagnetic system \cite{Dir}, where divergence is renormalized
into the electron mass.  They gave the renormalized equation of motion
for a global string comprised of the free part derived from the
Nambu-Goto action with the damping term, which becomes negligible for
a circular loop in the cosmological scale where $\ln(t/\delta) \sim
\CO(100)$. Then they concluded that the global string network can be
well approximated by the motion of the Nambu-Goto action. However, as
shown by Battye and Shellard \cite{BS1} though the calculation is done
in the flat spacetime, the kinks on long string are substantially
rounded due to the backreaction of NG boson radiation, which may
significantly affect the small scale structure (if at all) of the
global network system. Also, the above approach cannot include the
long-range force between long strings, which may decrease the energy
density of long strings. Thus, the examination of the dynamics of
global strings by use of the Kalb-Ramond action is not yet complete.

In the previous paper \cite{YKY}, we manipulated the equation of
motion of the complex scalar field instead of the Kalb-Ramond action.
Then, we showed that the large scale behavior of the global string
network goes into the scaling regime and $\xi$ for the global string
network becomes $\CO(1)$, which is significantly smaller than that for
the local string network.

In this paper, we give a comprehensive analysis of the evolution of
the global string network based on the model adopted in \cite{YKY}.
We show the scaling of the global string network under more general
situations and investigate the dependence of $\xi$ on boundary
conditions and some parameters. Then, the loop distribution function
is given in order to decide whether the small scale structure exists
like the local string network.

The paper is organized as follows: In the next section, we give the
formulation of numerical simulations. In the section 3, we give the
method of the identification of string segments and closed loops.
Then, the scaling parameter $\xi$ and the loop distribution function
are given. Finally, we discuss our results and give conclusions.

\section{Formulation of numerical simulations}

We consider the following Lagrangian density for a complex scalar
field $\Phi(x)$,
\beq
  \CL[\Phi] = g_{\mu\nu}(\del^{\mu}\Phi)(\del^{\nu}\Phi)^{\dagger}
                 - V_{\rm eff}[\Phi,T] \:,
\eeq 
where $g_{\mu\nu}$ is identified with the Robertson-Walker metric and
the effective potential $V_{\rm eff}[\Phi,T]$ is given by
\beq
  V_{\rm eff}[\Phi,T] = \frac{\lambda}{2}(\Phi\Phi^{\dagger} - \eta^2)^2 
                 + \frac{\lambda}{3}T^2\,\Phi\Phi^{\dagger} \:.
  \label{eqn:effpot}
\eeq

\begin{figure}[htb]
  \begin{center}
    \leavevmode\psfig{figure=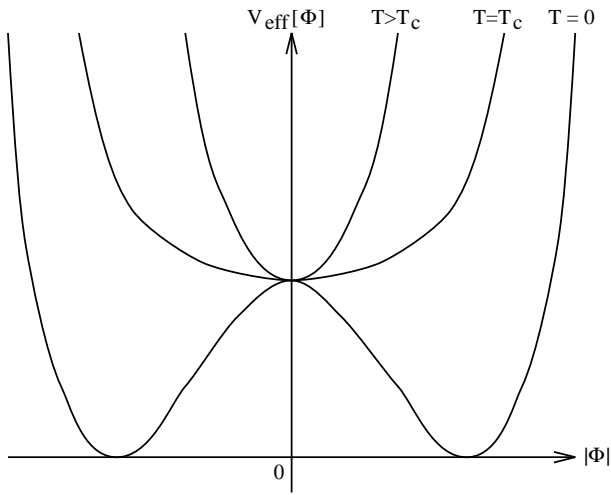,width=8cm}
  \end{center}
  \caption{One-loop finite temperature effective potential $V_{\rm
      eff}[\Phi,T]$ of the complex scalar field.}
  \label{fig:potential}
\end{figure}
For $T > T_{c} = \sqrt{3}\eta$, the potential $V_{\rm eff}[\Phi,T]$
has a minimum at $\Phi = 0$, and the $U(1)$ symmetry is restored. On
the other hand, new minima $|\Phi|_{\rm min} =
\eta\sqrt{1-(T/T_{c})^2}$ appear and the symmetry is broken for $T <
T_{c}$ (Fig.\ \ref{fig:potential}). In this case the phase transition
is of second order.

The equation of motion is given by
\beq
  \ddot{\Phi}(x) + 3H\dot{\Phi}(x) - \frac{1}{R(t)^2}\nabla^2\Phi(x)
   = - V'_{\rm eff}[\Phi,T] \:,
  \label{eqn:master}
\eeq
where the prime represents the derivative $\del/\del\Phi^{\dagger}$
and $R(t)$ is the scale factor. The Hubble parameter $H = \dot
R(t)/R(t)$ and the cosmic time $t$ are given by
\bea
  H^2 = \frac{8\pi}{3 m_{\rm pl}^2} \frac{\pi^2}{30} g_{*} T^4 \:,
   ~~~~~
  t = \frac{1}{2H} = \frac{\xi}{T^2} \:,
  \label{eqn:hubble}
\eea
where $m_{\rm pl} = 1.2 \times 10^{19}$GeV is the Plank mass, $g_{*}$
is the total number of degrees of freedom for the relativistic
particles, and radiation domination is assumed. We define the
dimensionless parameter $\zeta$ as
\beq
  \zeta \equiv \frac{\xi}{\eta}  = \lmk \frac{45M_{\rm pl}^2}
      {16\pi^3g_{*} \eta^2}
  \rmk^{1/2} \:.
  \label{eqn:zeta}
\eeq
In our simulation, we take $\zeta = 10$, which corresponds to $\eta
\sim (10^{15}-10^{16})$ GeV with $g_{\ast} = 1000$ and later
investigate $\zeta$ dependence on the result. The energy density at
each lattice point is written as
\beq
  \rho(x) = \dot\Phi(x) \dot\Phi^{\dagger}(x) 
           + \frac{1}{R(t)^2}\nabla\Phi(x) \cdot \nabla\Phi^{\dagger}(x) 
            + V_{\rm eff}[\Phi,T]  \:.
\eeq

We take the initial time $t_{i} = t_{c}/4$ and the final time $t_{f} =
75\,t_{i} = 18.75\,t_{c}$, where $t_c$ is the epoch $T=T_c$. Since the
$U(1)$ symmetry is restored at the initial time $t = t_{i}$, we adopt
as the initial condition the thermal equilibrium state with the mass
squared,
\beq
  m^2 = \left. \frac{d^2 V_{\rm eff}[|\Phi|,T]}{d|\Phi|^2} 
        \right|_{|\Phi|=0} \:,
\eeq
which is the inverse curvature of the potential at the origin at $t =
t_{i}$. In the thermal equilibrium state, $\Phi$ and $\dot\Phi$ are
Gaussian distributed with the correlation functions,
\bea
  \la \beta|\Phi(\vect x)\Phi^{\dagger}(\vect y)|\beta
                             \ra_{\rm equal\hbox{-}time} &=&
   \int \frac{d\vect k}{(2\pi)^3} \frac1{2\sqrt{\vect k^2 + m^2}}
           \coth{\frac{\beta\sqrt{\vect k^2 + m^2}}{2}}
             e^{i\vect k \cdot (\vect x-\vect y)} \:, \non  \\
             && \\
  \la \beta|\dot\Phi(\vect x)\dot\Phi^{\dagger}(\vect y)|\beta
                             \ra_{\rm equal\hbox{-}time} &=&
   \int \frac{d\vect k}{(2\pi)^3} \frac{\sqrt{\vect k^2 + m^2}}{2}
           \coth{\frac{\beta\sqrt{\vect k^2 + m^2}}{2}}
             e^{i\vect k \cdot (\vect x-\vect y)} \:. \non \\
             &&
\eea
The functions $\Phi(\vect x)$ and $\dot\Phi(\vect y)$ are uncorrelated
for $\vect x \ne \vect y$. We generate these fields for the initial
condition in the momentum space, because the corresponding fields
$\tilde{\Phi}(\vect k)$ and $\tilde{\dot\Phi}(\vect k)$ are
uncorrelated there. Then these fields are inverse Fourier transformed
into the configuration spaces by the FFT formalism.

Hereafter we measure the scalar field in units of $t_{i}^{-1}$, $t$
and $x$ in units of $t_{i}$, and the energy density in units of
$t_{i}^{-4}$. The equation of motion and the total energy density are
given by
\bea
  && \hspace{-1.8cm} 
   \ddot{\Phi}(x) + \frac{3}{2t}\dot{\Phi}(x) - \frac{1}{t}\nabla^2\Phi(x)
   = - \lmk |\Phi|^2 + \frac{\zeta^2}{36\,t} 
                                - \frac{\zeta^2}{144} \rmk 
    \Phi^{\dagger} \:, \\
  && \hspace{-1.8cm} 
   \rho(x) = \dot\Phi(x) \dot\Phi^{\dagger}(x) 
           + \frac{1}{t}\nabla\Phi(x) \cdot \nabla\Phi^{\dagger}(x) 
            + \frac12 \lmk |\Phi|^2 - \frac{\zeta^2}{144} \rmk^2
             + \frac{\zeta^2}{36\,t} |\Phi|^2 \:,
\eea
where $\lambda$ is set to unity for brevity. The scale factor $R(t)$
is normalized as $R(1) = 1$.
 
We perform the simulations in four different sets of lattice sizes and
spacings (See TABLE \ref{tab:set1}.): (1)~$128^3$ lattices with the
physical lattice spacing $\delta x_{\rm phys} = 2\sqrt{3}t_{i}
R(t)/25$.  (2)~$64^3$ lattices with $\delta x_{\rm phys} =
4\sqrt{3}t_{i} R(t)/25$. (3)~$256^3$ lattices with $\delta x_{\rm
  phys} = \sqrt{3}t_{i} R(t)/25$. (4)~$256^3$ lattices with $\delta
x_{\rm phys} = 2\sqrt{3}t_{i} R(t)/25$. In all cases, the time step is
taken as $\delta t = 0.01 t_{i}$. In the case (1), the box size is
nearly equal to the horizon volume $(H^{-1})^{3}$ and the lattice
spacing to a typical width $\delta \sim 1.0/(\sqrt{2}\eta)$ of a
string at the final time $t_{f}$. Furthermore, in order to investigate
the dependence of $\zeta$, we arrange the case with $\zeta = 5$,
(7)~$128^3$ lattices with the physical lattice spacing $\delta x_{\rm
  phys} = 2\sqrt{6}t_{i} R(t)/25$. In this case we follow the time
development of the system until the final time, $t_{f} = 150\,t_{i} =
37.5\,t_{c}$, when the box size is nearly equal to the horizon volume
$(H^{-1})^{3}$ and the lattice spacing to a typical width of a string.
For each case, we simulate the system from 10~((2), (3), (4), and (7))
or 300~((1)) different thermal initial conditions. Since the
simulation box is larger than the horizon volume even at the final
time of the simulation, we adopt the periodic boundary condition. But,
under the periodic boundary conditions, there exists no infinite
string so that it is possible that string completely disappears in the
simulation box. Therefore, in order to verify the dependence of
boundary conditions, we also simulate the cases in TABLE
\ref{tab:set2} under the reflective boundary condition where
$\nabla^{2}\Phi(x)$ on the boundary points disappears \footnote{Note
  that this condition is different from the open boundary condition
  where $\nabla\Phi(x)$ on the boundary points disappears. Under the
  open boundary condition, the string feels attractive forces from the
  boundary so that the number of the strings tends to decrease than
  that in the real universe as under the periodic boundary
  condition.}.
\begin{table}
\caption{Five different sets of the simulations under the periodic
  boundary condition}
\label{tab:set1}
  \begin{center}
     \begin{tabular}{cccccc}
        $$ & ${\rm lattice~number}$ &
        ${\rm lattice~spacing}$ &
        $\zeta$ & ${\rm realization}$ & $\xi$ \\
        $$ & $$ & $({\rm unit} = t_{i}R(t))$ & $$ & $$ & $$ \\
        \hline
        $(1)$ & $128^3$ & $2\sqrt{3}/25$ & $10$ & $300$ & $0.90\pm0.06$ \\
        $(2)$ & $~64^3$ & $4\sqrt{3}/25$ & $10$ & $10$ & $0.90\pm0.05$ \\
        $(3)$ & $256^3$ & $\sqrt{3}/25$  & $10$ & $10$ & $0.99\pm0.09$ \\
        $(4)$ & $256^3$ & $2\sqrt{3}/25$ & $10$ & $10$ & $0.97\pm0.04$ \\
        \hline \\
        \hline
        $(7)$ & $128^3$ & $2\sqrt{6}/25$ & $5$ & $10$ & $0.88\pm0.07$ \\
     \end{tabular}
  \end{center}
\end{table}
\begin{table}
\caption{Six different sets of the simulations under the reflective
  boundary condition}
\label{tab:set2}
  \begin{center}
     \begin{tabular}{cccccc}
        $$ & ${\rm lattice~number}$ &
        ${\rm lattice~spacing}$ &
        $\zeta$ & ${\rm realization}$ & $\xi$ \\
        $$ & $$ & $({\rm unit} = t_{i}R(t))$ & $$ & $$ & $$ \\
        \hline
        $(1)$ & $128^3$ & $2\sqrt{3}/25$ & $10$ & $10$ & $1.77\pm0.03$ \\
        $(2)$ & $~64^3$ & $4\sqrt{3}/25$ & $10$ & $10$ & $1.57\pm0.04$ \\
        $(3)$ & $256^3$ & $\sqrt{3}/25$  & $10$ & $10$ & $2.00\pm0.05$ \\
        $(4)$ & $256^3$ & $2\sqrt{3}/25$ & $10$ & $10$ & $1.30\pm0.02$ \\
        $(5)$ & $128^3$ & $4\sqrt{3}/25$ & $10$ & $10$ & $1.18\pm0.03$ \\
        $(6)$ & $256^3$ & $4\sqrt{3}/25$ & $10$ & $10$ & $1.03\pm0.03$ \\
     \end{tabular}
  \end{center}
\end{table}

Using the second order leap-frog method and the Crank-Nicholson
scheme, the discretized equation of motion reads
\bea
   \dot\Phi_{\vect i,n+1/2} &=& \frac{1}{1+\frac{3\delta t}{4t}}
       \lkk\,
          \lmk\,1-\frac{3\,\delta t}{4\,t}\,\rmk \dot\Phi_{\vect i,n-1/2}
            +\frac{\delta t}{t}\,\nabla^2\Phi_{\vect i,n} 
              - \delta t \lhk\,
        |\Phi_{\vect i,n}|^2 + \frac{\zeta^2}{36\,t} - \frac{\zeta^2}{144} 
                         \,\rhk \Phi_{\vect i,n}^2
       \,\rkk \:, \non \\
   \Phi_{\vect i,n+1} &=& \Phi_{\vect i,n}
         + \delta t\,\dot\Phi_{\vect i,n+1/2}
       \:, \non \\
   \nabla^2\Phi_{\vect i,n} &\equiv& \sum_{s = x, y, z}
       \frac{\Phi_{i_{s}+1_{s},n} - 2\,\Phi_{i_{s},n} +
             \Phi_{i_{s}-1_{s},n}}{(\delta x)^2} \:,
\eea
where $\vect i$ represents spatial index and $n$ temporal one.

\section{Results}

In order to judge whether the global string network relaxes into the
scaling regime, we give time development of $\xi$, which is defined as
\beq
   \rho = \xi \mu / t^2 \:,
\eeq
where $\mu \equiv 2\pi\eta^2\ln(t/(\delta\xi^{1/2}))$ is the average
energy per unit length of global strings.

\subsection{String identification}

\begin{figure}
    \begin{center}
      \leavevmode\psfig{figure=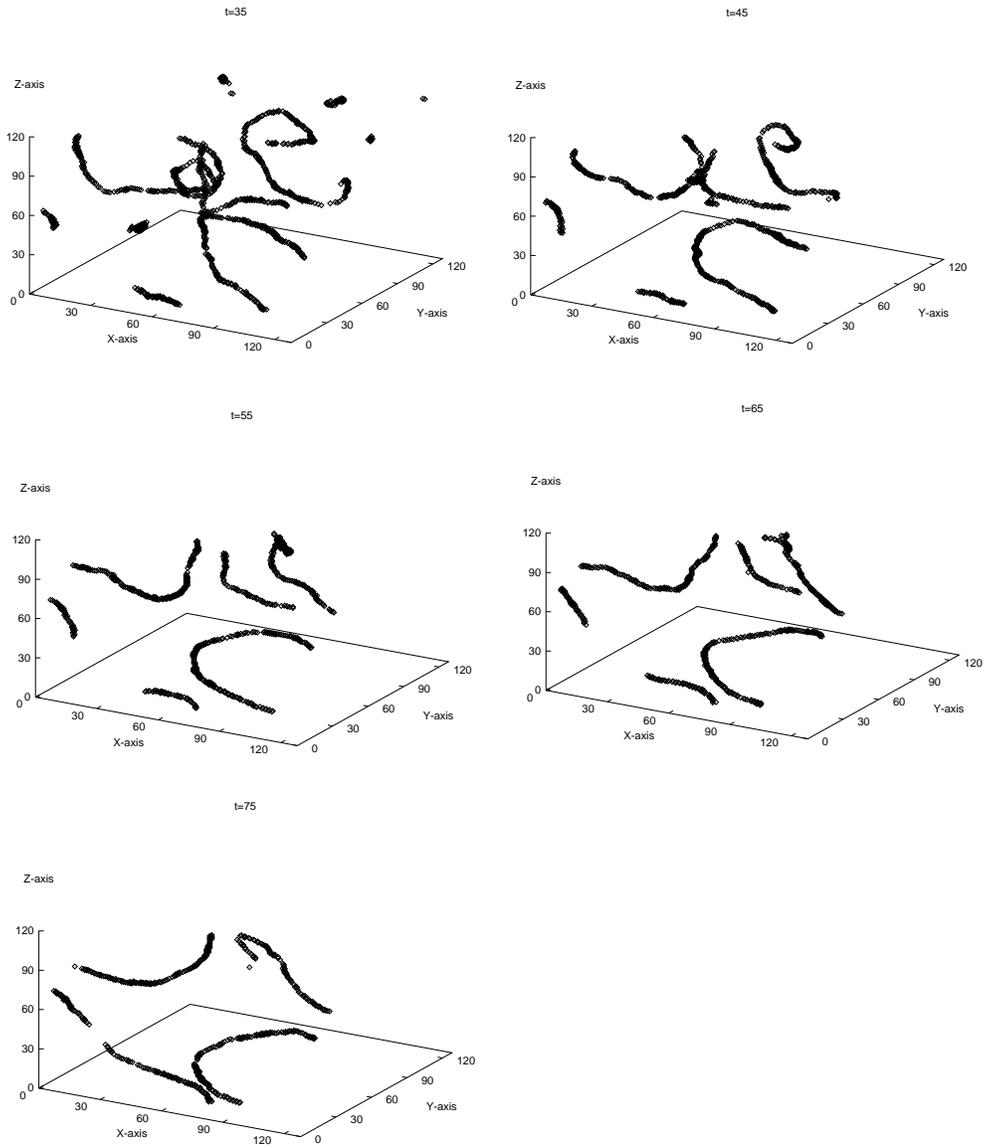,width=13cm}
    \end{center}
    \caption{Snapshots of a realization at $t = 35,45,55,65,75$ in the case
      (1) under the periodic boundary condition. Lattices identified
      with a part of a string core are shown.}
    \label{fig:stringpoint}
\end{figure}

Before obtaining $\xi$ and the loop distribution function, we must
identify the string segment. Since spacetime is discretized in our
simulations, a point with $\Phi = 0$ corresponding to a string core is
not necessarily situated at a lattice point. In the worst case, a
point with $\Phi = 0$ lies at the center of a plaquette.  Therefore,
we use a static cylindrically-symmetric solution, which is obtained by
solving the equation
\beq
  \frac{d^2 f}{dr^2} + \frac{1}{r}\frac{df}{dr}
   - \frac{f}{r^2} - V'_{\rm eff}[f,T] = 0 \:,
\eeq
with $\Phi(r, \theta) \equiv f(r)e^{i\theta}$ and the winding number
$n = 1$. The boundary conditions are given by
\bea
  f(r) &\rightarrow& |\Phi|_{\rm min}, \qquad (r \rightarrow \infty) \:, \\
  f(0) &=& 0\:.
\eea
We require that a lattice is identified with a part of a string core
if the potential energy density there is larger than that
corresponding to the field value of a static cylindrically-symmetric
solution at $r = \delta x_{\rm{phys}}/\sqrt{2}$. Then only one lattice
within a section of a straight string core is identified with a string
segment except the case where a point with $\Phi = 0$ lies at the
center of a plaquette.  Of course, the real string is not exactly
straight but bent and more complex. But, as seen in Fig.\ 
\ref{fig:stringpoint}, our identification is good and only one lattice
within a section of a string core is identified with a part of a
string core. In order to solve the above equation of motion, we have
to use the standard shooting technique, which is the repeated process
so that it is bothersome to follow the above procedure for each time
step. Instead, we obtain the solutions every $500\,\delta t$ and make
the fitting formula. For intermediate time steps, we judge whether a
lattice point belongs to the string segment by comparing the potential
energy obtained from the simulation and that from the fitting formula.
Thus, by counting the number of the lattices identified with a part of
a string core, we can evaluate the total length of strings within the
horizon volume, $(2t)^3$, from which the energy density can be
obtained.

\subsection{Scaling property}

\begin{figure}
  \begin{center}
    \leavevmode\psfig{figure=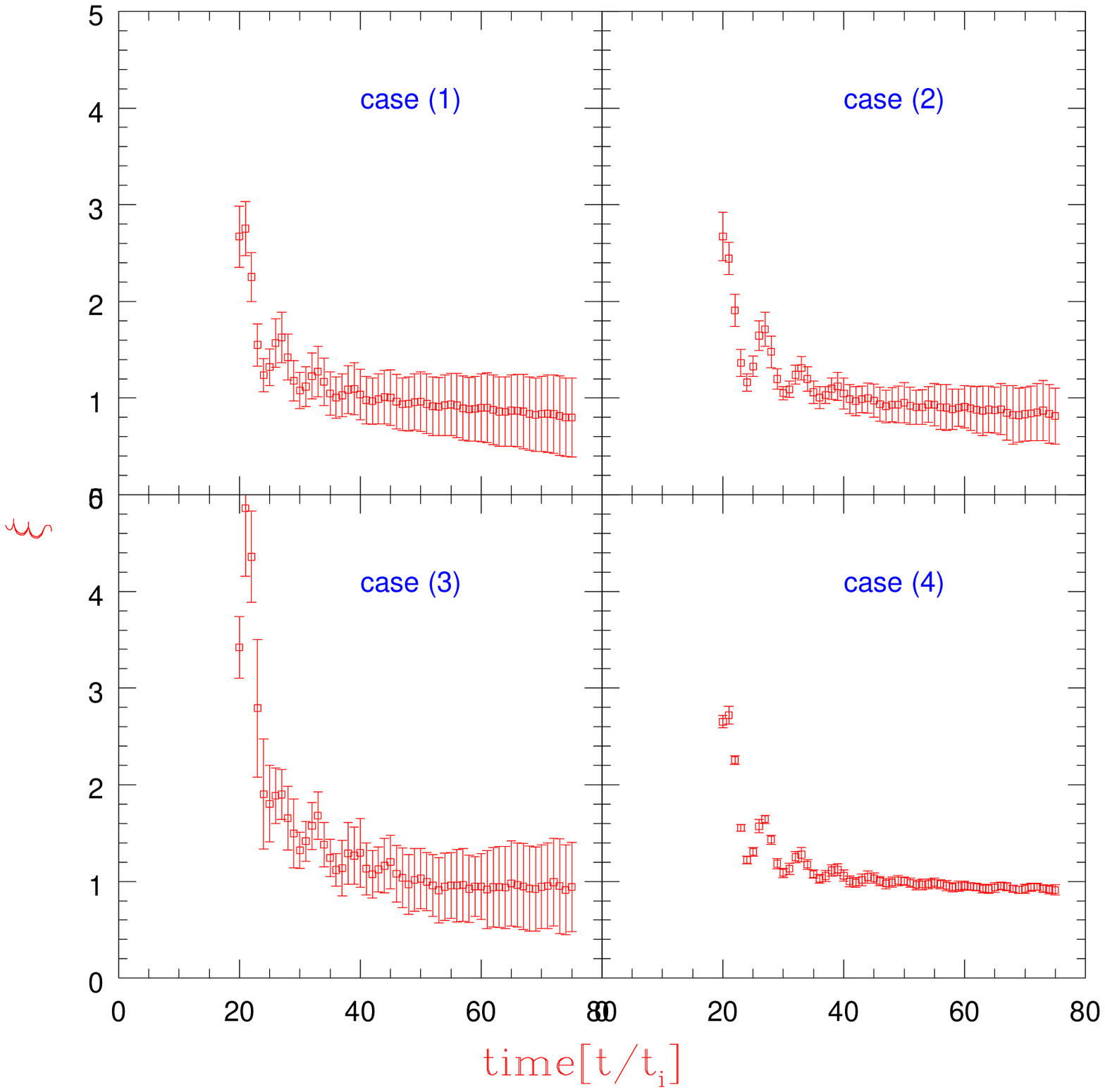,width=15cm}
  \end{center}
  \caption{Time development of $\xi$ in the cases from (1) to
    (4) under the periodic boundary condition. Symbols($\Box$)
    represent time development of $\xi$.  The vertical lines denote a
    standard deviation over different initial conditions.}
  \label{fig:xi}
\end{figure}

\begin{figure}
  \begin{center}
    \leavevmode\psfig{figure=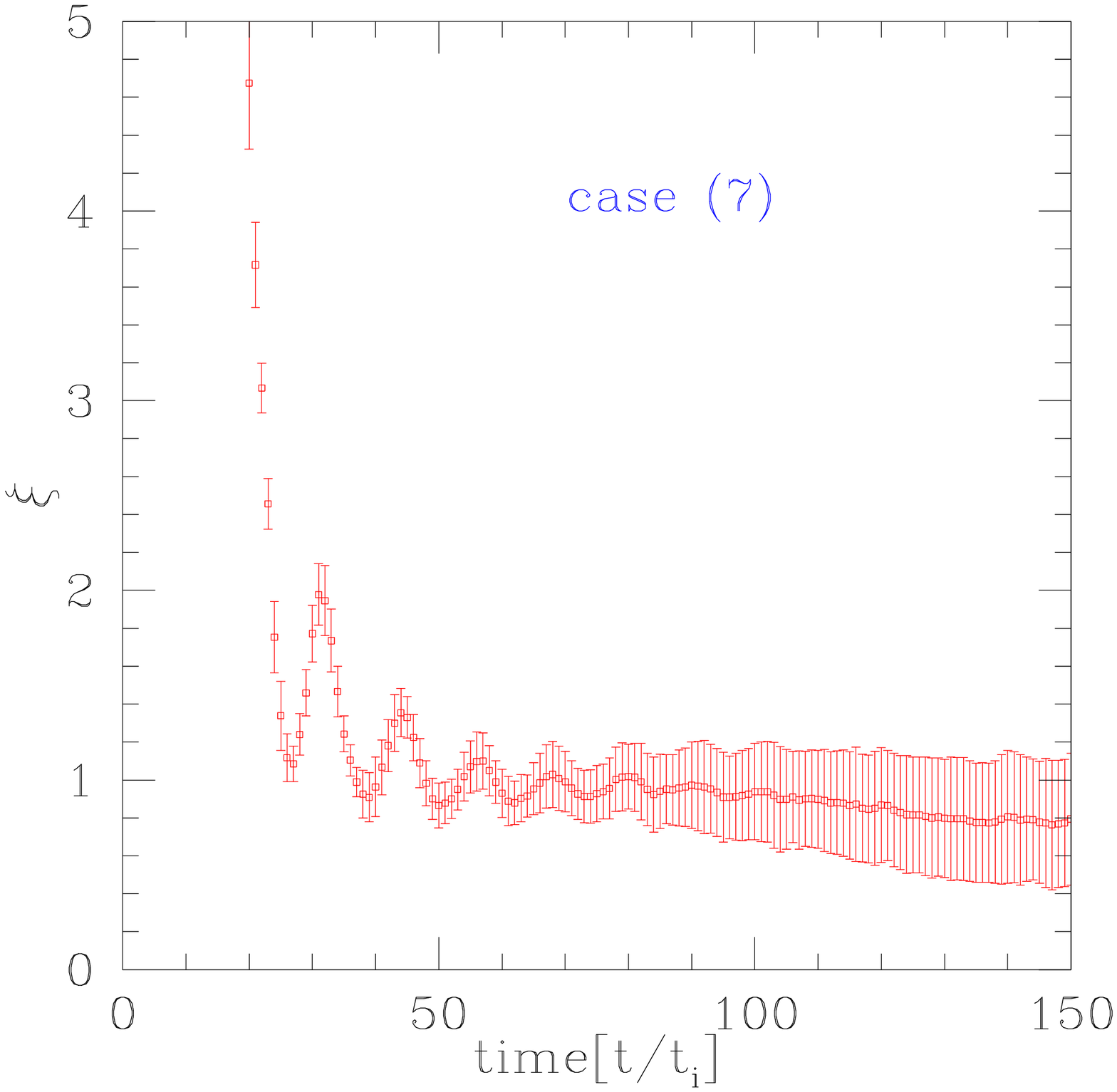,width=8cm}
  \end{center}
  \caption{Time development of $\xi$ in the case (7). Symbols($\Box$)
  represent time development of $\xi$. The vertical lines denote a
  standard deviation.}
  \label{fig:xi2}
\end{figure}

Time development of $\xi$ in the cases from (1) to (4) is described in
Fig. \ref{fig:xi}. We find that after some relaxation period, $\xi$
becomes a constant irrespective of time with (1)$0.99\pm0.09$,
(2)$0.97\pm0.04$, (3)$0.90\pm0.06$, and (4)$0.90\pm0.05$. They all are
consistent within the standard deviation. Thus we can conclude that a
global string network relaxes into scaling regime in the radiation
domination. We also show time development of $\xi$ in the case (7) in
Fig. \ref{fig:xi2}. Then $\xi$ asymptotically becomes a constant,
$0.88\pm0.07$, which is also consistent with the above all cases with
$\zeta = 10$ within the standard deviation. Hence we can also conclude
that $\zeta$ does not change the essential result. Note that the
standard deviation in the case (2) is much smaller than the other
cases because the box in the case (2) includes more horizon volumes at
each time. Also, $\xi$ seems to oscillate at the early epoch because
the homogeneous mode of the field oscillates in the radial direction
of the potential, which rapidly decays. In fact, the period of the
oscillation coincides with $2\pi$ times the inverse mass at the
potential minimum.

\begin{figure}
  \begin{center}
    \leavevmode\psfig{figure=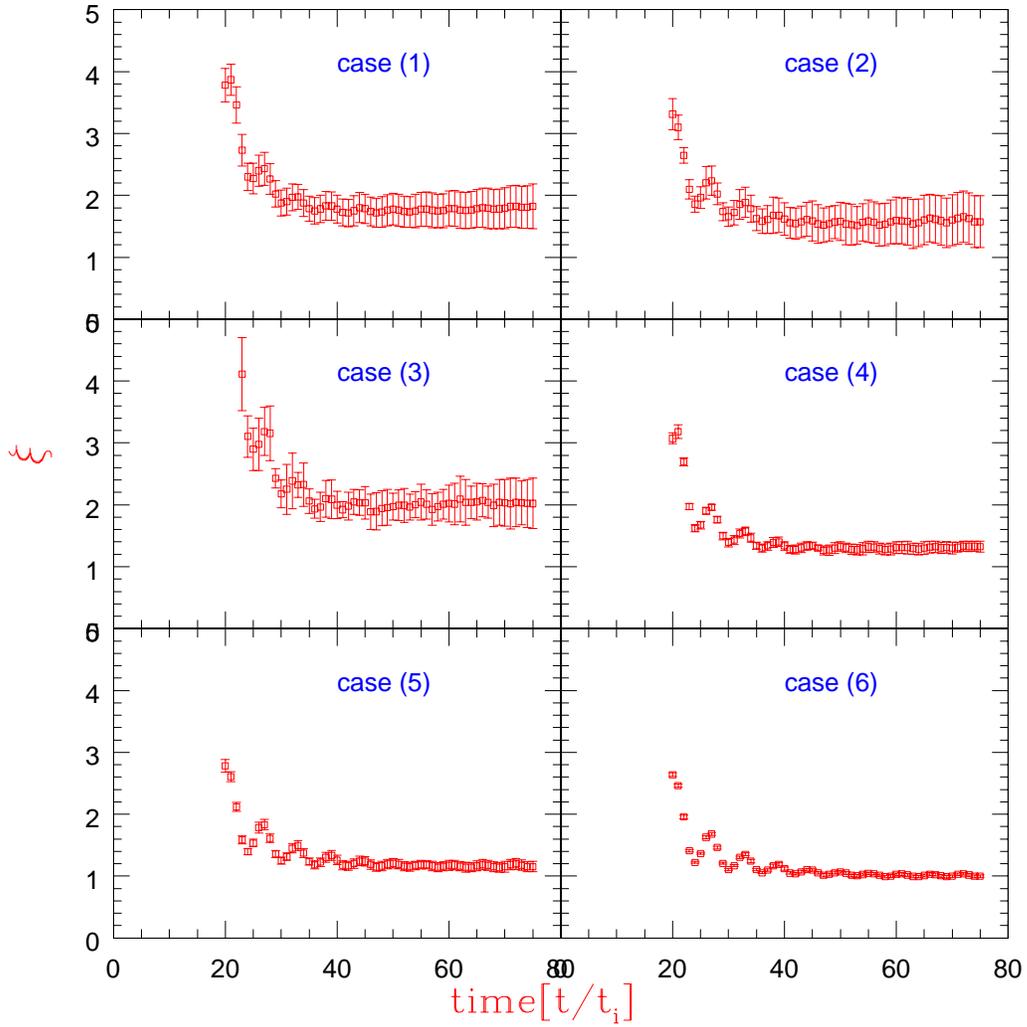,width=15cm}
  \end{center}
  \caption{Time development of $\xi$ in the cases under the reflective
    boundary condition. Symbols($\Box$) represent time development of
    $\xi$. The vertical lines denote a standard deviation.}
  \label{fig:xi3}
\end{figure}

Fig. \ref{fig:xi3} represents the results under the reflective
boundary condition, where $\xi$ becomes a constant irrespective of
time with (1)$1.77\pm0.03$, (2)$1.57\pm0.04$, (3)$2.00\pm0.05$,
(4)$1.30\pm0.02$, (5)$1.18\pm0.03$, and (6)$1.03\pm0.03$. Though all
results are consistent with that under the periodic boundary condition
within the factor two, the former tends to become larger than the
latter. This is because under the reflective boundary condition,
strings are repulsed by the boundary and the string near the boundary
intercommutes less often than that near the center of the simulation
box because the partner to intercommute only lies in the inner
direction of the boundary. The results in large box simulations,
(4)-(6), are converging so that it is safe to say that if we take the
box size larger than $2^{3}$ times the horizon volume, the reflective
boundary condition has no effect on the results.

\subsection{Loop distribution}

We also investigate the loop distribution. Since in our simulations it
is judged by the potential energy whether a lattice point is a part of
the string, the string is not necessarily continuous. Therefore, we
identify a closed loop as follows; First we select a lattice point
belonging to the string segment. Then we connect it with the nearest
neighbor among lattice points belonging to the string segment. We
proceed this process one after another until the connection returns to
the starting lattice point.

\begin{figure}
    \begin{center}
      \leavevmode\psfig{figure=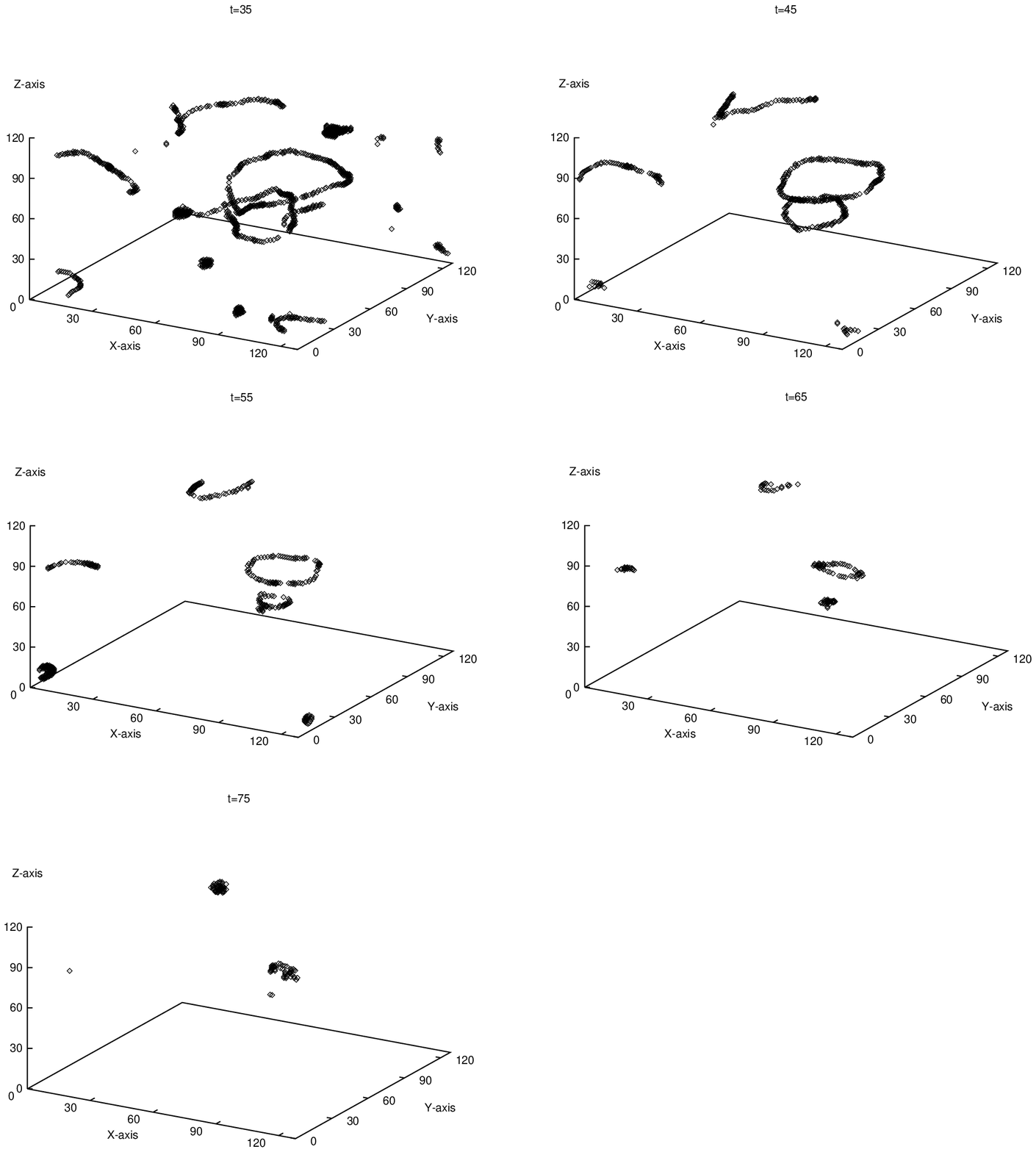,width=13cm}
    \end{center}
    \caption{Snapshots of a realization at $t = 35,45,55,65,75$ in the case
      (1) under the periodic boundary condition. Loops decay through
      radiating NG bosons.}
    \label{fig:decay}
\end{figure}

Kibble's one scale model predicts the loop distribution as
\beq 
  n_{l}(t) = \frac{\nu}
     {t^{\frac32} (l + \kappa t)^{\frac52}} \:,
\eeq
where $\nu$ is a constant, $l$ is the length of a loop, and the
log-dependence of $\mu$ is neglected. Different from a local string,
the dominant energy loss mechanism for global strings is radiation of
the associated Nambu-Goldstone field \cite{GB}. We define the
radiation power, $P$, as $P = \kappa \mu$ where $\kappa$ is a
constant. An example of decay of a closed loop is depicted in Fig.
\ref{fig:decay}.

\begin{figure}
  \begin{center}
    \leavevmode\psfig{figure=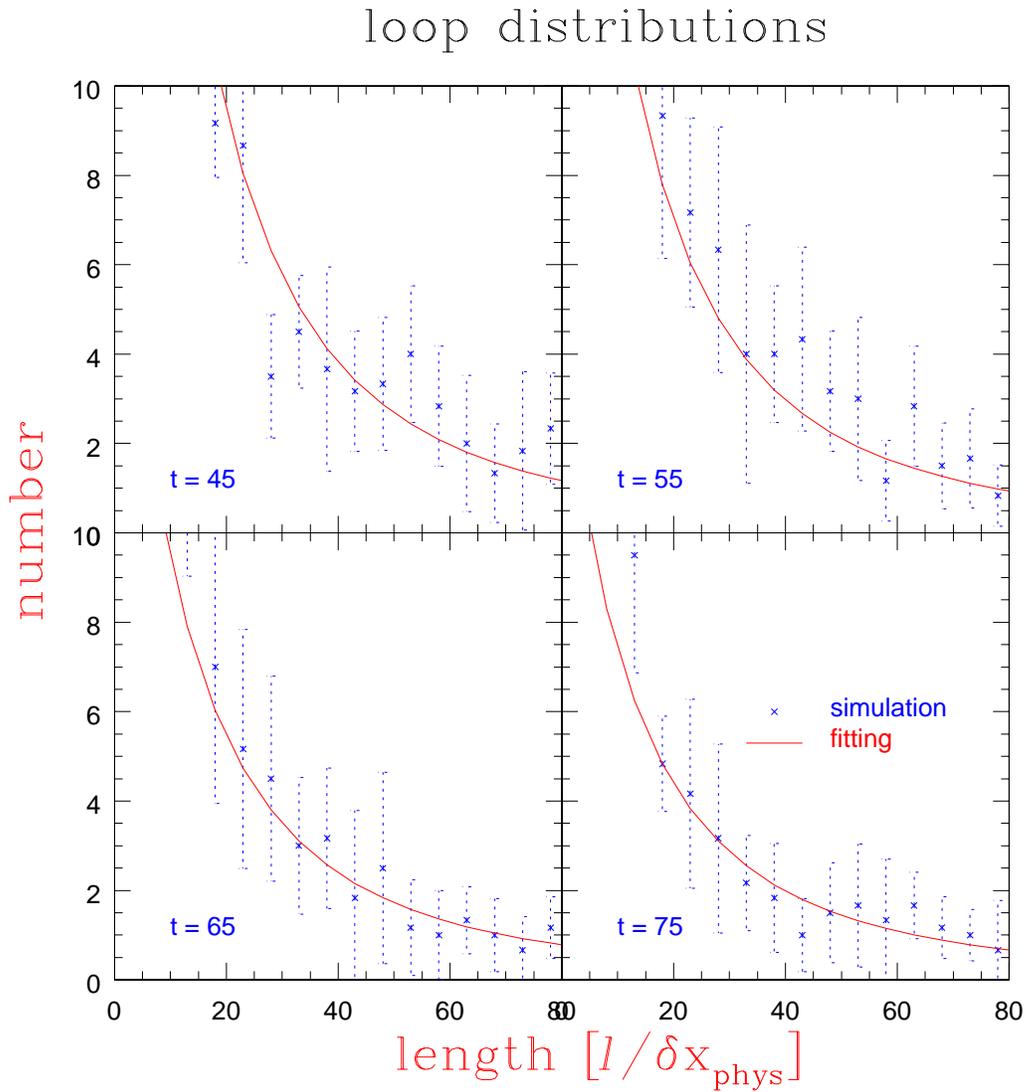,width=15cm}
  \end{center}
  \caption{Loop distributions at $t = 45, 55, 65, 75$ are
    depicted. The number is summed over the box size ($128(\delta
    x)^3$) and 50 realizations for each groups. Bins are cut every
    5$\times\delta x$.}
  \label{fig:loop}
\end{figure}

\begin{figure}
  \begin{center}
    \leavevmode\psfig{figure=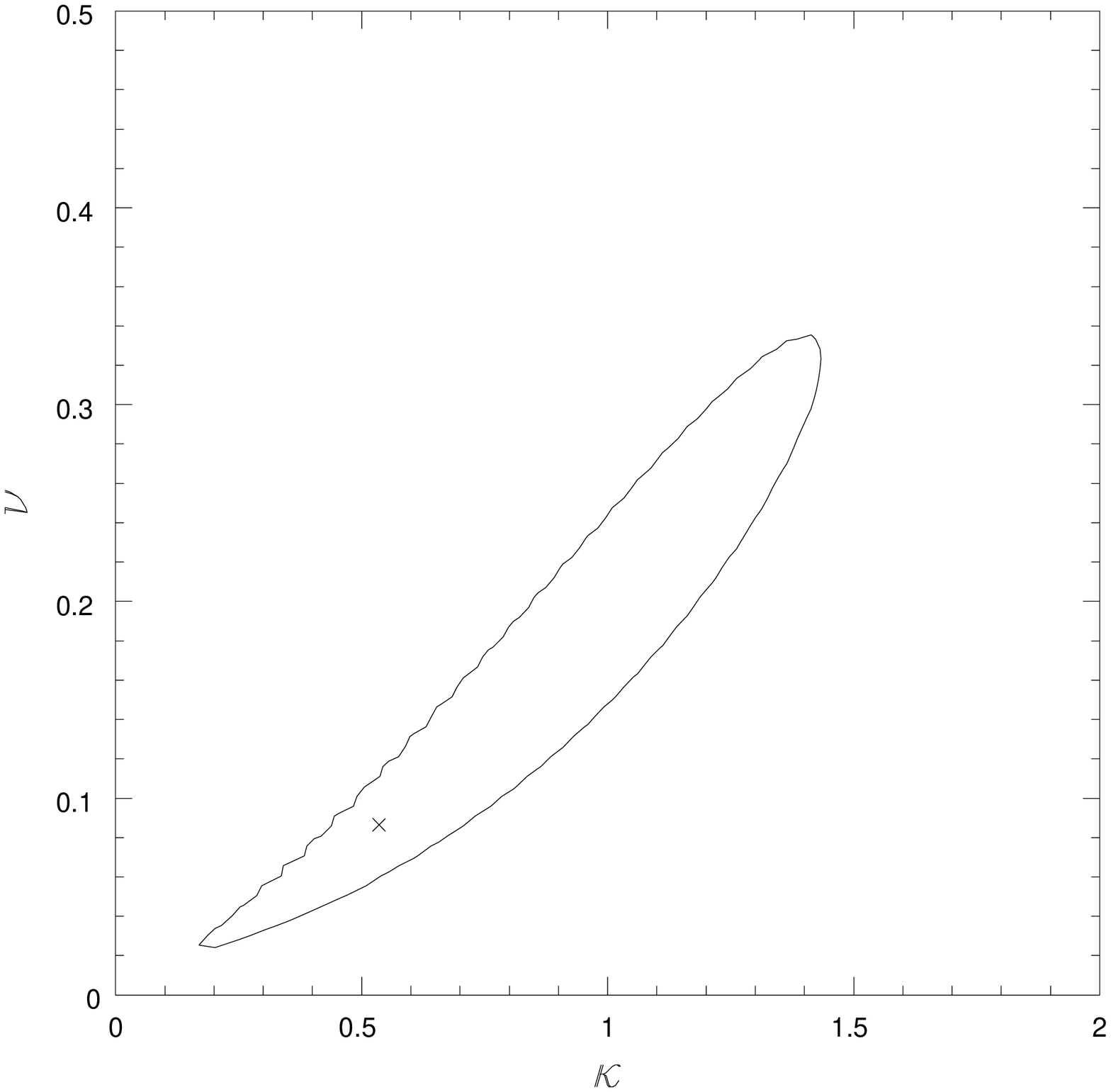,width=8cm}
  \end{center}
  \caption{Fittings for $\kappa$ and $\nu$. The cross point represents the
    best fit values for $\kappa$ and $\nu$. The solid circle denotes
    $68\%$ C.L.}
  \label{fig:kappanu}
\end{figure}

We determine whether the loop distributions in the simulation
coincides with the above predicted loop distribution function.  The
loop distributions in the case (1) at different times ($t = 45, 55,
65, 75 t_{i}$) are described in Fig. \ref{fig:loop}. Since long
strings are rare, we cut the length of loops into bins with the width
5$\times\delta x$. Also, we divide 300 realizations into 6 groups
comprised of 50 realizations and we summed the number of loops over 50
realizations for each groups. The dot represents the number of loops
averaged over 6 groups and the dash line represents the standard
deviation. They can be simultaneously fitted with the above formula if
one takes $\kappa \sim 0.535$ and $\nu \sim 0.0865$. Fittings for
$\kappa$ and $\nu$ are also given in Fig. \ref{fig:kappanu}. Thus, the
loop production function as well as the large scale behavior of the
string scales together for the global string network.

As an another evidence for the scaling of the loop production, we
consider the NG boson spectrum radiated from strings. If loops are
formed at scales much smaller than the horizon distance, there should
be significant power of radiated NG bosons for modes corresponding to
the scale at which loops are formed. For the purpose, we first
represent the complex field $\Phi(t, \vect x)$ in terms of the radial
mode $\chi(t, \vect x)$ and the NG boson field $\alpha(t, \vect x)$ as
\beq
  \Phi(t,\vect x) = \lkk \eta + \frac{\chi(t,\vect x)}{\sqrt2} \rkk
               \exp \lmk \frac{i\alpha(t, \vect x)}{\sqrt2 \eta} \rmk \:.
  \label{phasedef}
\eeq
Then the kinetic energy density of NG bosons is given by
\bea
  \frac12 \dot\alpha(t, \vect x)^2 &=&
      \frac{\eta^2}{|\Phi(t, \vect x)|^4}
   \non \\
   && \hspace{-1.5cm}
    \times
     \lkk - {\rm Im}\Phi(t, \vect x) {\rm Re}\dot\Phi(t, \vect x)
          + {\rm Re}\Phi(t, \vect x) {\rm Im}\dot\Phi(t, \vect x)
     \rkk^2 \:.
\eea
One may wonder if the power spectrum obtained by the Fourier
transformation of the above kinetic energy density is what we want.
But it includes the contribution from NG bosons formed at the symmetry
breaking as well as that radiated from strings. Also, the
decomposition of the field into the radial and phase modes is
inadequate for the lattice point near the string segment.

Then, we evaluate the average energy density of NG bosons radiated in
the period between $t_{1}$ and $t_{2}$, $\bar\rho[t_{1},t_{2}]$.  For
the purpose, we subtract the redshifted kinetic energy at $t_{1}$ from
the kinetic energy at $t_{2}$ since emitted NG bosons damp like
radiation. Thus, $\bar\rho[t_{1},t_{2}]$ is given by
\bea
  \bar{\rho}[t_{1},t_{2}] &=&
  \frac{1}{V} \int d^3\vect x \rho[t_{1},t_{2};\vect x] \non \\*
                          &=&
       \frac{1}{V} \int d^3\vect x \lkk
           \frac12 \dot\alpha(t_{2}, \vect x)^2 -
           \frac12 \dot\alpha(t_{1}, \vect x)^2
                                   \lmk
           \frac{t_{1}}{t_{2}}
              \rmk^2 \rkk
  \non \\
                            &=&
       \frac{1}{V} \int \frac{d^3\vect k}{(2\pi)^3}
       \lkk \frac12\, |\dot\alpha_{\vect k}(t_{2})|^2 -
            \frac12\, |\dot\alpha_{\vect k}(t_{1})|^2 \lmk
            \frac{t_{1}}{t_{2}} \rmk^2 \rkk
  \non \\
                            &\equiv&
  \int \frac{d^3\vect k}{(2\pi)^3}
         \tilde{\rho}_{\vect k}[t_{1},t_{2}]
                             \equiv
  \int_{0}^{\infty} \frac{dk}{2\pi^2} {\rho_{k}}[t_{1},t_{2}]
\:,
\label{eqn:spectrum}
\eea
where $V$ is the simulation volume and $\dot\alpha(t, \vect x) = \int
\frac{d^3\vect k}{(2\pi)^3} \dot{\alpha}_{\vect k}(t) \exp(i \vect k
\cdot \vect x)$.

Furthermore, to avoid contamination of string cores to the spectrum of
emitted axions, we divide the simulation box into 8 cells and stock
the field data of a cell if there are no string cores in that cell
between $t_{1}$ and $t_{2}$. Over all such cells, we average power
spectra of kinetic energy of axions obtained through Fourier
transformation. We follow the above procedure between $t_{1} =
65t_{i}$ and $t_{2} = 75t_{i}$ for the case (3) under the periodic
boundary condition, which is the highest resolution simulation, but
with the zero temperature potential $V_{\rm eff}[\Phi, T = 0]$ after
$t = 20 t_{i}$ because the decomposition of the field becomes
well-defined. One may suspect that this spectrum is dominated by the
kinetic energy of NG bosons associated with global strings (string NG
bosons) rather than that of free NG bosons radiated from global
strings. But this is incorrect for the following reason; The total
energy of string NG bosons is almost as much as that of free NG
bosons. However, the contribution to the energy of string NG bosons is
dominated by the gradient energy because the kinetic energy of string
NG bosons is much smaller that their gradient energy by the factor
$v^2$ ($v$ is the velocity of the string core and $v \ll 1$ except
just before the disappearance of the loop).  On the other hand, the
gradient and the kinetic energy of free NG bosons is the same.
Therefore the kinetic energy of free NG bosons is much larger than
that of string NG bosons, that is, our spectrum is dominated by the
former.  Also, even if it contributed, the kinetic energy of string NG
bosons would decay in proportion to $t^{-2}$, so that its contribution
has been removed from the spectrum in our method as done in Eq.
(\ref{eqn:spectrum}).

\begin{figure}
  \begin{center}
    \leavevmode\psfig{figure=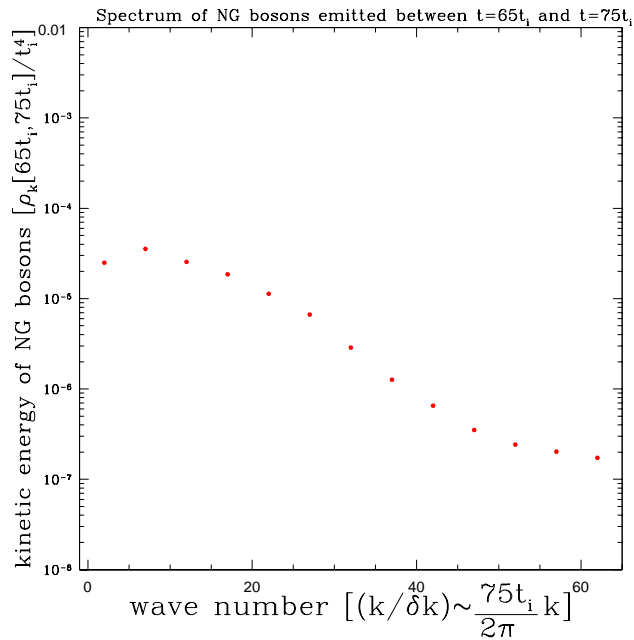,width=8.6cm}
  \end{center}
  \caption{Filled dots represent the power spectrum, $\rho_k$, which
    is already averaged over the direction of $\vect{k}$ and
    multiplied by the phase-space factor as defined in
    Eq.(\ref{eqn:spectrum}). Bins are cut every $5\delta k$. $k =
    64\delta k$ corresponds to string cores.}
  \label{fig:spectrum}
\end{figure}

The result is depicted in Fig.\ \ref{fig:spectrum}, which has already
been given in \cite{YKY} in the different context. The spectrum of
emitted NG bosons is highly peaked at the horizon scale. Thus, loops
are formed not at much smaller scale than the horizon distance but
around the horizon scale.

\section{Discussions and conclusions}

In this paper we gave a comprehensive investigation on the evolution
of the global string network in the radiation dominated universe by
use of the numerical simulation based on the complex scalar field
model which spontaneously breaks the U(1) symmetry.

In order to decide whether the large scale behavior of the global
string network goes into the scaling regime, we followed time
development of the scaling parameter $\xi$ which characterizes the
average energy density of global strings. We found that $\xi$ is
almost constant irrespective of cosmic time under both the periodic
and the reflective boundary condition. All the results are consistent
within the factor two. But $\xi$ under the reflective boundary
condition tend to become larger than that under the periodic boundary
condition.  This can be understood as follows; Under the periodic
boundary condition, there are no infinite strings and strings with two
boundary points on the opposite planes always intercommute with the
partner so that $\xi$ tends to become small. On the other hand, under
the reflective boundary condition, strings are repulsed by the
boundary. Furthermore the string near the boundary intercommutes less
often than that near the center of the simulation box because the
partner to intercommute only lies in the inner direction of the
boundary. Thus, $\xi$ tends to become large.  Therefore, $\xi$ in the
real world should lie between that under the periodic boundary
condition and that under the reflective boundary condition.
Considering all the cases under the periodic boundary condition and
the cases (4)-(6) of the large box simulations under the reflective
boundary condition, it is safe to say that $\xi \sim (0.9-1.3)$.

You should note that $\xi$ is much smaller than that of a local
string, which is of the order of ten. This is mainly because global
strings can intercommute more often than local strings since an
attractive force proportional to the inverse separation works between
a global string and a global anti-string.

We have also investigated the loop distribution. It can be well fitted
to that predicted by the one scale model if we take $\nu \sim 0.0865$
and $\kappa \sim 0.535$. Thus, the loop production grows with the
horizon distance and we did not observe the small scale structure.
This is because the Nambu-Goldstone(NG) boson radiation from strings
is so efficient that the small scale structures on strings are damped
out. The damping scale is typically $\kappa t$, which is near the
horizon distance in contrast with $G\mu t \ll t$ (G : the gravitational
constant) if at all for the local string network. Bennett \cite{BEN}
showed that unless produced loops with the length of the horizon
distance often self-intersect and fragment into smaller loops with the
typical length smaller than the horizon distance, the reconnection
rate is large enough to prevent scaling. In the global string network,
parent loops do not fragment into smaller loops with the typical
length smaller than the horizon distance, instead they rapidly shrink
through radiating NG bosons so that the reconnection rate of parent
large loops becomes small and scaling should be kept.

\subsection*{Acknowledgments}
The author is grateful to Jun'ichi Yokoyama and Masahiro Kawasaki for
useful discussions. This work was partially supported by the Japanese
Grant-in-Aid for Scientific Research from the Monbusho, Nos.\ 
10-04558.

\end{document}